\documentclass[twocolumn]{aastex631}

\usepackage{apjfonts}
\usepackage{times}
\usepackage{CJK}
\usepackage{amsmath}
\usepackage{hyperref}
\usepackage{cleveref}
\usepackage{graphicx}
\usepackage{natbib}
\usepackage{hyperref}
\usepackage{amssymb}
\usepackage{rotating}
\usepackage{lipsum} 
\usepackage{academicons}
\usepackage[utf8]{inputenc}

\usepackage{color}

\shorttitle{}
\shortauthors{Le et al.}

\newcommand{\kms}{km~s$^{\rm -1}$}
\newcommand{\ergs}{erg s$^{-1}$}
\def\gsim{\mathrel{\rlap{\lower4pt\hbox{\hskip1pt$\sim$}}
    \raise1pt\hbox{$>$}}}         

\def\lsim{\mathrel{\rlap{\lower4pt\hbox{\hskip1pt$\sim$}}
    \raise1pt\hbox{$<$}}}         

\newcommand{\ustc}{\affil{CAS Key Laboratory for Research in Galaxies and Cosmology, Department of Astronomy, University of Science and Technology of China, Hefei 230026, China; \href{mailto: lha@ustc.edu.cn}{lha@ustc.edu.cn}; \href{mailto:qinc@mail.ustc.edu.cn}{qinc@mail.ustc.edu.cn}; \href{mailto: xuey@ustc.edu.cn}{xuey@ustc.edu.cn}}}
\newcommand{\sustc}{\affil{School of Astronomy and Space Science, University of Science and Technology of China, Hefei 230026, China}}
\newcommand{\hcm}{\affil{Faculty of Physics and Engineering Physics, University of Science, Ho Chi Minh City, Vietnam}}
\newcommand{\hcmvn}{\affil{Vietnam National University, Ho Chi Minh City, Vietnam}}

\begin{document}

\begin{CJK*}{UTF8}{gbsn}

\title{Active Galactic Nuclei and STaR fOrmation in Nearby Galaxies (AGNSTRONG). I. Sample and Strategy}

\author[0000-0003-1270-9802]{Huynh Anh N. Le} \ustc \sustc
\author[0009-0000-0126-8701]{Chen Qin} \ustc \sustc
\author[0000-0002-1935-8104]{Yongquan Xue} \ustc \sustc
\author[0000-0002-1653-4969]{Shifu Zhu} \ustc \sustc
\author[0000-0003-4410-8737]{Kim Ngan N. Nguyen} \hcm \hcmvn
\author[0009-0005-3916-1455]{Ruisong Xia} \ustc \sustc
\author[0000-0002-4926-1362]{Xiaozhi Lin} \ustc \sustc


\begin{abstract} 

We introduce our project, AGNSTRONG (Active Galactic Nuclei and STaR fOrmation in Nearby Galaxies). Our research goals encompass investigating the kinematic properties of ionized and molecular gas outflows, understanding the impact of AGN feedback, and exploring the coevolution dynamics between AGN strength activity and star formation activity. We aim to conduct a thorough analysis to determine whether there is an increase or suppression in SFRs among targets with and without powerful relativistic jets. Our sample consists of 35 nearby AGNs with and without powerful relativistic jet detections. Utilizing sub-millimeter (sub-mm) continuum observations at 450 $\rm \mu m$ and 850 $\rm \mu m$ from SCUBA-2 at the James Clerk Maxwell Telescope, we determine star-formation rates (SFRs) for our sources using spectral energy distribution (SED) fitting models. Additionally, we employ high-quality, spatially resolved spectra from UV-optical to near-infrared bands obtained with the Double Spectrograph and Triple Spectrograph mounted on the 200-inch Hale telescope at Palomar Observatory to study their multiphase gas outflow properties. This paper presents an overview of our sample selection methodology, research strategy, and initial results of our project. We find that the SFRs determined without including the sub-mm data in the SED fitting are overestimated by $\sim$0.08 dex compared to those estimated with the inclusion of sub-mm data. Additionally, we compare the estimated SFRs in our work with those traced by the 4000\AA\ break, as provided by the MPA-JHU catalog. We find that our determined SFRs are systematically higher than those traced by the 4000\AA\ break. Finally, we outline our future research plans.

\end{abstract}



\section{Introduction}\label{sec:intro}



The mechanisms triggering active galactic nucleus (AGN) activities and regulating star formation (SF) processes are often referred to as AGN feedback. Many theoretical studies postulate that AGN feedback plays an important role in coevolution between supermassive black holes (SMBHs) and their host galaxies \citep[e.g.,][]{Croton+06}. Despite considerable efforts invested in theoretical modeling and observational exploration, our comprehension of the intricate relationship between AGN activity strength and SF processes within host galaxies remains controversial. This controversy is highlighted by diverse observational outcomes \citep[e.g.,][]{Ferrarese+00, Woo+06, Xue+10, Alexander+12, Kormendy+13, Le+14, Xue+17}.

In the context of recent observational results, instances of both negative and positive AGN feedback have been identified in individual AGNs and within large statistical samples. Diverse observational findings are attributed, in part, to the challenges associated with quantifying AGN strength (including black hole mass and accretion rate), star-formation rates (SFRs), and the constraints imposed by sample selection \citep[e.g.,][]{Silk+98, Le+17a, Le+17b, Zhuang+20, Shin+19}. These results underscore the complexity of AGN feedback dynamics and the interactions between AGN activity and the surrounding interstellar medium (ISM). Further observational endeavors are necessary to unravel the underlying physical intricacies of the interplay between AGN strength and SF activities.

One of the significant challenges in the study of AGN feedback is the difficulty of estimating SFRs \citep[e.g.,][]{Kennicutt+98, Kennicutt+09, Leroy+2012, Whitcomb+2023}. SFR estimation is a complex endeavor with considerable uncertainties, often reaching approximately 0.7 dex \citep{Salim+16}. These uncertainties are closely tied to the choice of SFR indicators, which encompass various wavelength ranges, including rest-frame ultraviolet (UV), optical, and infrared (IR) bands.

The UV diagnostic, exemplified by features such as the 4000\AA\ break, offers a direct measure of SFR originating from young stellar populations. Nevertheless, this approach is hampered by dust obscuration, introducing significant uncertainties and limiting its ability to account for dust-obscured star formation \citep{Salim+07}. On the contrary, optical emission lines (e.g., [O II] 3727\AA, H$\alpha$), when corrected for dust obscuration, are commonly employed for SFR estimation, especially in the context of low-redshift AGNs \citep{Zhuang+19}. However, this method is not without limitations, as it can be susceptible to contamination from the ionizing radiation of ISM shocks and/or AGN emissions. 

In the IR band diagnostic, there may be contributions from the dusty torus heated by the AGN, particularly evident in the mid-infrared (MIR) emission \citep[e.g.,][]{Ramos+17, Garcia+22}. The far-infrared (FIR) diagnostic stands out as a robust SFR indicator since it predominantly traces the cold dust emission from star formation, minimally affected by ISM shocks or AGN-related dusty torus emission \citep[e.g.,][]{Ellison+16a, Lopez+18, Fuller+19, Xu+20}. To obtain reliable SFR estimates, a meticulous decomposition of dust heated by the AGN is essential when using spectral energy distribution (SED) fitting. Observations in the FIR and/or sub-millimeter (sub-mm) wavelengths are indispensable for securing accurate SFRs because they are necessary for reliably tracing the peak of interstellar dust emission at $\sim$100 $\rm \mu m$ in SED fitting models. In particular, for targets lacking observed FIR fluxes, the inclusion of sub-mm fluxes assumes an important role within SED fitting models. Recent findings by \citet{Kim+22} underscore the critical nature of incorporating FIR/sub-mm fluxes in SED fitting models to avoid an overestimation of IR luminosity by a factor of 2.

As sub-mm observations play a crucial role in determining SFRs in AGNs, we have carefully selected a local sample of AGNs for the purpose of observing their continuum fluxes at 450 $\rm \mu m$ and 850 $\rm \mu m$ wavelengths using SCUBA-2 at the James Clerk Maxwell Telescope (JCMT). Using this sample, our primary dedication lies in accurately determining the SFRs and subsequently investigating the relationships between AGN strength and SF activities. Our project centers on the precise determination of SFRs and the study of AGN feedback, encompassing both quasar- (QSO-) and radio-mode feedback. Our overarching goal is to explore the interplay between AGN activity levels and the rates of star formation within their host galaxies. In addition to utilizing JCMT data, we employ data from various spectrographs to facilitate comprehensive analysis, including the Double spectrograph (DBSP) and Triple spectrograph (TPSP) mounted on the 200-inch Hale (P200) telescope at Palomar Observatory and the Prime Focus Spectrograph (PFS, \citealp{Takada+2014}) to be mounted on the 8.2-m Subaru telescope at Mauna Kea Observatory. 

In Section \ref{section:sample}, we describe the sample selection process. Section \ref{section:analysis} provides detailed information on the measurements and analysis plans. Our plans with the PFS is shown in Section \ref{section:pfs}. Section \ref{section:result} presents the initial results and our future plans. The summary is provided in Section \ref{section:sum}. Similar to \citet{Brinchmann+04}, we adopt the \citet{Kroupa01} stellar initial mass function (IMF) in this work. The following cosmological parameters are used throughout the paper: $H_0 = 70$~km~s$^{-1}$~Mpc$^{-1}$, $\Omega_{\rm m} = 0.30$, and $\Omega_{\Lambda} = 0.70$.

\section{Sample Selection and Observations}\label{section:sample}


\subsection{Sample Selection} \label{sub:sample}

The sample, characterized by the presence of detected gas outflows in their emission line spectra and corresponding radio counterparts, presents a unique opportunity for testing and investigating AGN feedback in the local universe. \citet{Rakshit+18} utilized the `specObj' data products classified as `QSO' from the Sloan Digital Sky Survey (SDSS) DR12 catalog \citep{Alam+15}. They identified that there are $\sim$5,000 type-1 AGNs with signal-to-noise ratios (S/N) $>$ 10 at the 5100 \AA\ continuum and S/N $>$ 5 of H$\beta$ at redshift $z < 0.3$. 

\begin{figure*}
\centering
\includegraphics[width = 0.45\textwidth]{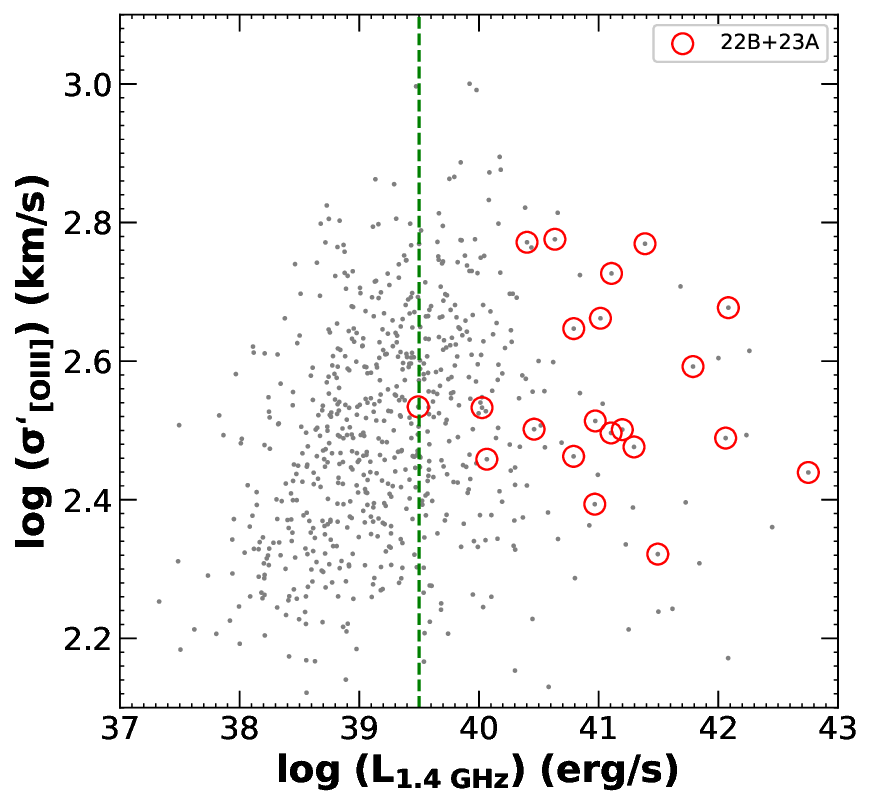}
\includegraphics[width = 0.456\textwidth]{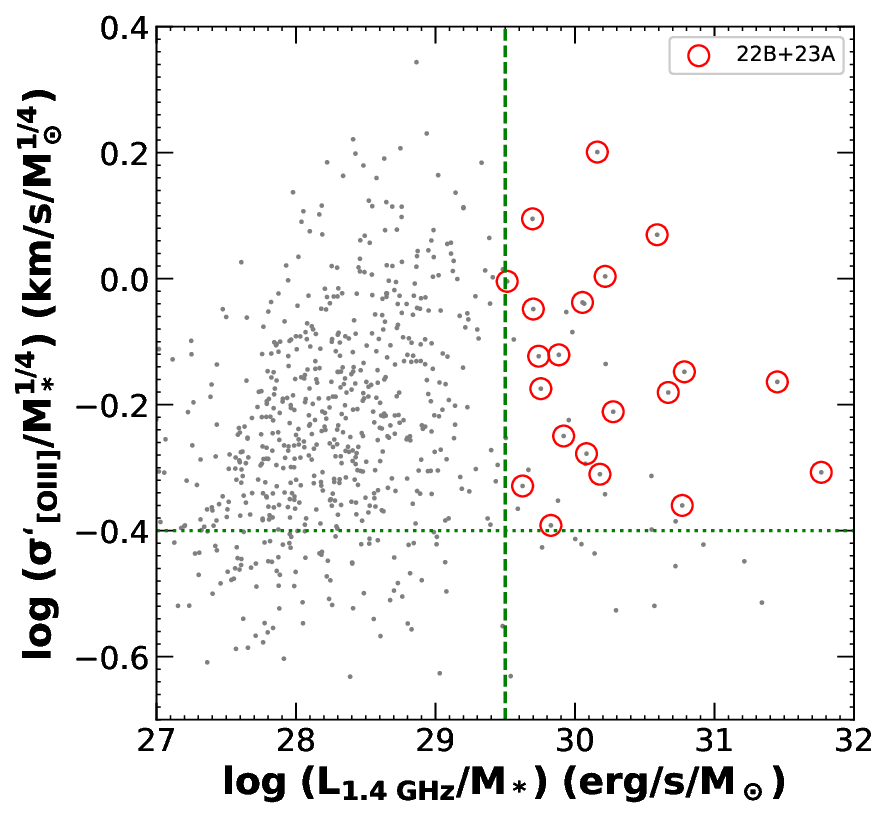}
\caption{Selection of AGNs (z $<$ 0.3) with powerful outflows and high radio luminosities from the SDSS type-1 AGNs \citep{Rakshit+18}. Grey dots indicate the SDSS sample. Points with red circles represent our 21 selected objects in the proposals M22BP083 and M23AP023, respectively. Left panel: Distribution of outflow velocity $\sigma_{\rm [OIII]}'$ as a function of radio luminosity. Right panel: Distribution of log $\rm \sigma_{\rm [OIII]}'/M_{*}^{1/4}$ as a function of radio luminosity normalized by stellar mass. Note that M$_{*}^{1/4}$ serves as a proxy for stellar velocity dispersion. The vertical dashed line and horizontal dotted line denote $\rm L_{1.4 GHz}$/M$_*$ $>$ 10$^{29.5}$ erg s$^{-1}$ and log $\sigma_{\rm [OIII]}'$/M$_*$$^{1/4} \gtrsim$ $-$0.4 (indicating powerful radio jets and outflows), respectively. The grey dots without red circles within the threshold green lines on the top-right portion of the right panel denote data available in the JCMT archive. \label{fig:agnradio_select}}
\end{figure*}

By cross-matching these sources, selected from the SDSS, with the FIRST VLA Survey at 1.4 GHz, we found 918 AGNs that exhibit radio flux densities (integrated flux of the entire jet) above 1 mJy. Due to the limited time available to propose to the JCMT under the `Expanding Partner Program'  (less than 14 hours per semester), from these 918 AGNs, we selected 21 AGNs (Figure \ref{fig:agnradio_select}) with high radio luminosity and strong outflow kinematics (L$_{\rm1.4 GHz}$ $>$ 10$^{29.5}$ erg s$^{-1}$ and log $\sigma_{\rm [OIII]}'$/M$_{*}$ $>$ $\sim$$-$0.4). In this selection, we defined $\sigma_{\rm [OIII]}'$ as a representation of the outflow velocity, considering the effects of inclinations, geometry of the outflows, extinction, turbulent velocity, etc., where $\sigma_{\rm [OIII]}'^{2}$ $=$ $\sigma_{\rm [OIII]}^{2}$ + $\rm V_{[OIII]}^{2}$. In this equation, $\sigma_{\rm [OIII]}$ and $\rm V_{\rm [OIII]}$ are velocity dispersion and velocity shift of the emission line, respectively. 

\begin{figure*}
\centering
\includegraphics[width = 0.43\textwidth]{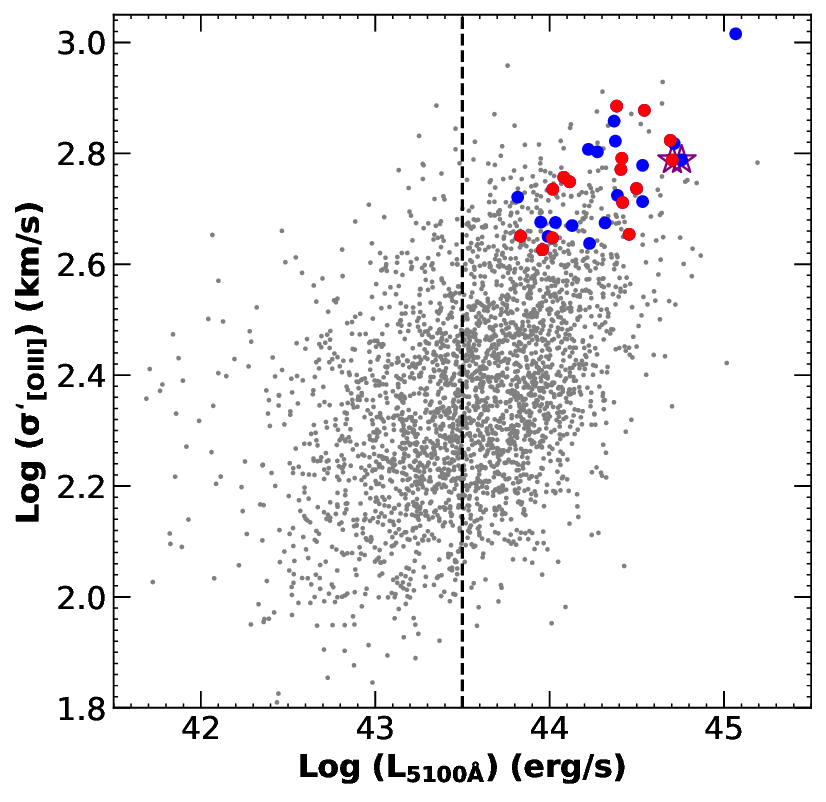}
\includegraphics[width = 0.86\textwidth]{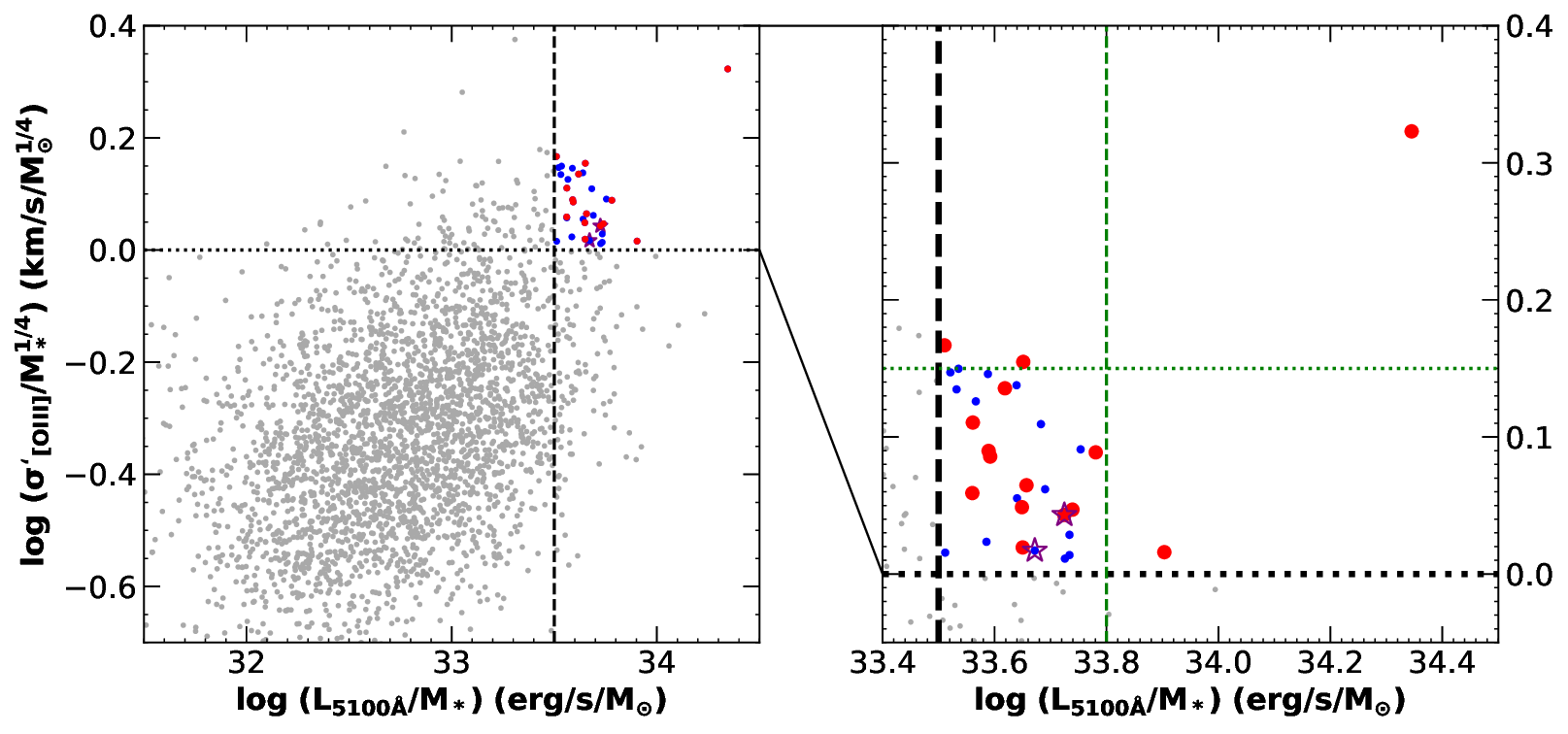}
\caption{Selection of AGNs (z $<$ 0.3) with powerful outflows and very high optical continuum luminosity from the SDSS Type-1 AGNs \citep{Rakshit+18}. Gray dots indicate the SDSS sample. Top panel: Distribution of log $\mathrm{\sigma_{[OIII]}'}$ as a function of optical continuum luminosity at $\mathrm{5100\text{\AA}}$. The black dash line indicates $\mathrm{L_{5100\text{\AA}}\ >\ 10^{43.5}\ erg\ s^{-1}}$. Bottom left panel: The distribution of log $\mathrm{\sigma_{[OIII]}'/M_*^{1/4}}$ is displayed as a function of $\mathrm{L_{5100\text{\AA}}}$ normalized by stellar mass ($\rm M_{*}$; from the MPA-JHU catalog). Note that $\mathrm{M_{*}^{1/4}}$ is a proxy for stellar velocity dispersion. The black dashed and dotted lines represent $\mathrm{L_{5100\text{\AA}}/M_{*}\ >\ 10^{33.5}}$ and $\mathrm{log\ \sigma'_{[OIII]}/M_{*}\ >\ 0}$ (indicating high luminosity and powerful outflows), respectively. There are 32 targets (red and blue points) that satisfy these criteria. Purple stars indicate targets that have been observed and are available in the SCUBA-2 sub-mm archival data. Bottom right panel: A zoomed-in view of the left panel for a detailed look at our selected sample (red points). Interestingly, one of our selected targets exhibits extremely high continuum luminosity and outflow strength properties ($\mathrm{L_{5100\text{\AA}}/M_{*}\ >\ 10^{33.8}}$ and $\mathrm{log\ \sigma'_{[OIII]}/M_{*}\ >\ 0.15}$, represented by green dashed and dotted lines, respectively).\label{fig:agn_select}}
\end{figure*}

Additionally, among the parent SDSS sample of $\sim$5,000 type-1 AGNs \citep{Rakshit+18}, we have also identified around 3,000 AGNs without jets with reliable [O III] line detection. Among these objects, we have observed that more luminous ones tend to exhibit higher [O III] velocity dispersion (as shown in Figure \ref{fig:agn_select}). These relatively luminous sources with the kinematic signature of ionized gas outflows present an intriguing opportunity to study the connection between AGN and SF activities in their host galaxies among the AGNs without jets. We have selected 14 targets with extreme properties, such as $\mathrm{L_{5100\text{\AA}}\ >\ 10^{43.5}\ erg\ s^{-1}}$, $\mathrm{L_{5100\text{\AA}}/M_{*}>10^{33.8}\ erg\ s^{-1}\ M_{\odot}^{-1} }$ or $\mathrm{log\ \sigma'_{[OIII]}/M_{*} > 0.15}$, as well as those with the lowest redshift ($\mathrm{z\leq 0.231}$). 

In short, we have selected 35 AGNs, 21 with powerful relativistic jets and 14 without, to study the coevolution dynamics between AGN activity and star formation. Table \ref{table:1} shows the physical properties of our selected sample. Typical root mean square (rms) values for the observations are within the range of 50$-$250 mJy at 450 $\rm \mu m$ and 2$-$10 mJy at 850 $\rm \mu m$, based on $\sim$1 hour of integration time per observation.

\subsection{Observations and Data Reduction}

\subsubsection{Observation facilities}


We use 450 $\rm \mu m$ and 850 $\rm \mu m$ observations using the SCUBA-2 on the JCMT as part of the `Expanding Partner Program' (IDs: M22BP038, M23AP023, and M23BP013,  PI: Kim Ngan Nhat Nguyen). 21 hours are allocated for observing 21 Type-1 AGNs with powerful relativistic jets and strong [O III] dispersion. All 21 targets have already been observed in the 2022B and the 2023A semesters. Additionally, the targets without jet detections have been scheduled for observations with the JCMT/SCUBA-2 in the 2023B semester, where we have been allocated 30 hours with band 4.

In addition, some of the targets in our sample have UV, optical, and near-infrared (NIR) spectroscopic observations using the DBSP and TPSP at P200 through the Telescope Access Program (TAP) of National Astronomical Observatories, Chinese Acadamy of Sciences (CAS). The targets consist of 11 nearby AGNs, observed with TPSP (ID: CTAP2022-A0037, PI: Huynh Anh N. Le) and DBSP (IDs: CTAP2022-A0050 and CTAP2022-B0058, PI: Yibo Wang). The targets are characterized by strong outflow velocities, with S/N $>$ 10 at the 5100\AA\ continuum and S/N > 5 of H$\beta$. Among these sources, five targets exhibit high radio luminosities, as detected by the VLA FIRST Survey at 1.4 GHz ($\rm L_{1.4 GHz} > 10^{41}\ erg\ s^{-1}$). These AGNs display strong outflow signatures in the ionized gas kinematics, with velocity shifts of up to 400 $\rm km\ s^{-1}$ and dispersions reaching 600 $\rm km\ s^{-1}$.

The long-slit spectroscopic observations conducted with TPSP and DBSP provide us with spatially-resolved measurements over scales spanning several kiloparsecs along the slit directions of TPSP (30") and DBSP (120"). This enables us to examine the kinematics of molecular gas in relation to that of the ionized gas. The TPSP offers high sensitivity, excellent spatial and spectral resolution, and comprehensive wavelength coverage in the J, H, and K-bands (1.0 $-$ 2.4 $\rm \mu m$). Similarly, the DBSP covers the UV-optical bands (3,000 $-$ 10,000 \AA). These attributes, combined with their use on the P200 telescope, make TPSP and DBSP an ideal setup for measuring spatially resolved flux and kinematics of molecular and ionized gas emission lines.

\subsubsection{Data Reduction of the JCMT data}

To reduce the observed JCMT data, we employ the SCUBA-2 pipeline ORAC-DR \citep{Jenness+15}, which utilizes the Starlink software \citep{Currie+14} and is based on the configuration of ${dimmconfig\_blank\_field.lis}$. We process the scan data with the default of matched filter sizes of 20" for 450 $\rm \mu m$ and 30" for 850 $\rm \mu m$. Subsequently, to obtain flux measurements at 450 $\rm \mu m$ and 850 $\rm \mu m$, we utilize the PICARD package to crop the images into 180" radius circles centered on the target locations. To assess the detected fluxes from the SCUBA-2 observations, we categorize the flux as either a detection or an upper limit. We compare the cropped images, overplotted with 3$\sigma$ sub-mm contours, with optical images sourced from the Digitized Sky Survey (DSS) at the European Southern Observatory (ESO). If a 3$\sigma$ sub-mm contour is situated at the center of the DSS image, we classify it as a detection. Furthermore, if there is a $>2\sigma$ clump in the sub-mm contour at the center of the DSS image location, we apply Gaussian convolution to smooth the data with three times the effective FWHM derived from a two-component fit of the beams, i.e., 25.8 and 37.8 at ${\rm 450\ \mu m}$ and ${\rm850\ \mu m}$, respectively \citep{Mairs+21}. If the clump exceeds $3\sigma$ of the convoluted data, we classify the flux as detected. Conversely, if no $>3\sigma$ clump is present, we classify the detection as an upper limit. In cases where the images display point sources, we measure the flux using the peak value position at the center of the target. If the targets exhibit extended structures, we employ the FELL-WALKER algorithm within the CUPID package \citep{Berry+17} to measure the flux of the image. The flux measurements are in units of mJy. The detailed process for SCUBA-2 data reduction is described in the JCMT tutorial\footnote{https://www.eaobservatory.org/jcmt/science/reductionanalysis-tutorials/}.

\subsubsection{Data Reduction of the TPSP and DBSP data}

The data reduction process for TPSP and DBSP observations entails several detailed steps facilitated by PypeIt \citep{Prochaska+2020}, a Python package specifically tailored for semi-automated astronomical and spectroscopic data reduction. Notably, PypeIt accommodates over 20 spectrographs, including TPSP and DBSP. This process encompasses essential procedures such as flat correction, wavelength calibration, cosmic ray removal, flux calibration, co-adding of exposures, and telluric correction. For flux calibration, we utilized the same standard stars observed on the same night as our targets. From the resulting flux-calibrated 2D spectra, we extract 1D spectra with aperture sizes ranging from 1" to 2". Utilizing these spatially resolved 1D spectra, we study the kinematics of gas and stars as a function of radius along the slit. Our goal is to investigate any potential connection between gas kinematics and radio jets in our sources.


\begin{deluxetable*}{ccccccccc}
\centering
\tabletypesize{\footnotesize}
\tablecaption{Physical properties of the selected sources\label{table:1}}
\tablewidth{0pt}
\tablehead{
\colhead{Observation ID} & \colhead{Name} & \colhead{RA} & \colhead{Dec} & \colhead{Redshift} & \colhead{$\rm T_{int}$ (hrs)} & \colhead{r-band mag} & \colhead{Morphology} & \colhead{Radio jets}
}
\startdata
(1) & (2) & (3) & (4) & (5) & (6) & (7) & (8) & (9) \\
\hline 
23BID01 & J0749+4510 & 07:49:06.50 & +45:10:33.8 & 0.192 & 1.000 & 16.68 & Uncertain & Y \\
23BID02 & J0754+3547 & 07:54:44.08 & +35:47:12.7 & 0.257 & 1.000 & 18.51 & Elliptical & Y \\
23BID03 & J1119+3151 & 11:19:02.26 & +31:51:22.5 & 0.262 & 1.000 & 18.10 & Elliptical & Y \\
23BID04 & J1244+4051 & 12:44:19.96 & +40:51:36.8 & 0.249 & 1.000 & 17.57 & Elliptical & Y \\
23BID05 & J1449+4221 & 14:49:20.71 & +42:21:01.2 & 0.179 & 1.000 & 17.28 & Merging & Y \\
23BID06 & J1527+2233 & 15:27:57.67 & +22:33:04.0 & 0.254 & 1.000 & 16.63 & Uncertain & Y \\
23BID07 & J1643+3048 & 16:43:31.91 & +30:48:35.5 & 0.184 & 1.000 & 17.11 & Uncertain & Y \\
23AID01 & J2351$-$0109 & 23:51:56.16 & $-$01:09:13.4 & 0.174 & 1.061 & 15.53 & Uncertain & Y \\
23AID02 & J1105+0202 & 11:05:38.88 & +02:02:57.4 & 0.106 & 1.006 & 16.03 & Uncertain & Y \\
23AID03 & J1345+5332 & 13:45:45.36 & +53:32:52.4 & 0.136 & 1.176 & 16.86 & Elliptical & Y \\
23AID04 & J1557+4507 & 15:57:07.20 & +45:07:00.1 & 0.234 & 1.064 & 18.38 & Uncertain & Y \\
23AID05 & J0936+3921 & 09:36:21.60 & +39:21:32.0 & 0.210 & 1.064 & 17.82 & Uncertain & Y \\
23AID06 & J1153+5831 & 11:53:24.00 & +58:31:38.3 & 0.202 & 1.176 & 17.80 & Uncertain & Y \\
23AID07 & J1140+4622 & 11:40:48.00 & +46:22:04.8 & 0.115 & 1.064 & 15.63 & Uncertain & Y \\
23AID08 & J0945+3521 & 09:45:25.92 & +35:21:03.6 & 0.208 & 1.008 & 17.63 & Uncertain & Y \\
23AID09 & J1532+0453 & 15:32:28.80 & +04:53:58.4 & 0.218 & 1.061 & 17.99 & Uncertain & Y \\
23AID10 & J1030+3102 & 10:30:59.04 & +31:02:55.7 & 0.178 & 1.008 & 16.48 & Uncertain & Y \\
23AID11 & J1312+3515 & 13:12:17.76 & +35:15:20.9 & 0.235 & 1.008 & 15.51 & Uncertain & Y \\
23AID12 & J1012+1638 & 10:12:56.16 & +16:38:53.2 & 0.118 & 1.006 & 17.59 & Uncertain & Y \\
23AID13 & J1005+3414 & 10:05:07.92 & +34:14:24.4 & 0.162 & 1.008 & 16.72 & Merging & Y \\
23AID14 & J1027+2859 & 10:27:36.48 & +28:59:17.5 & 0.257 & 1.008 & 19.05 & Uncertain & Y \\
23BID01 & J2220+0109 & 22:20:24.59 & +01:09:31.3 & 0.213 & 1.061 & 16.54 & Merging & N \\
23BID02 & J2123$-$0828 & 21:23:47.83 & $-$08:28:42.9 & 0.218 & 1.171 & 17.81 & Uncertain & N \\
23BID03 & J2318+1349 & 23:18:08.41 & +13:49:16.1 & 0.256 & 1.006 & 18.40 & Merging & N \\
23BID04 & J1004+0513 & 10:04:20.13 & +05:13:00.4 & 0.160 & 1.006 & 16.36 & Uncertain & N \\
23BID05 & J1313+5421 & 13:13:08.66 & +54:21:15.4 & 0.297 & 1.176 & 17.85 & Uncertain & N \\
23BID06 & J0835+0553 & 08:35:53.46 & +05:53:17.1 & 0.204 & 1.006 & 16.94 & Uncertain & N \\
23BID07 & J0900+3354 & 09:00:45.29 & +33:54:22.3 & 0.227 & 1.008 & 17.94 & Elliptical & N \\
23BID08 & J1029+4042 & 10:29:12.58 & +40:42:19.7 & 0.147 & 1.064 & 17.62 & Elliptical & N \\
23BID09 & J1553+0951 & 15:53:28.49 & +09:51:02.1 & 0.192 & 1.006 & 16.61 & Uncertain & N \\
23BID10 & J1616+1627 & 16:16:59.11 & +16:27:22.5 & 0.216 & 0.990 & 17.97 & Uncertain & N \\
23BID11 & J1123+2849 & 11:23:18.08 & +28:49:45.6 & 0.300 & 1.008 & 17.69 & Uncertain & N \\
23BID12 & J1041+2828 & 10:41:11.97 & +28:28:05.1 & 0.211 & 1.008 & 16.37 & Uncertain & N \\
23BID13 & J1632+0954 & 16:32:34.16 & +09:54:00.3 & 0.218 & 1.006 & 18.17 & Elliptical & N \\
23BID14 & J1421+1520 & 14:21:38.50 & +15:20:45.8 & 0.231 & 0.990 & 17.27 & Elliptical & N \\
\enddata
\tablecomments{Col. (1): Observational identification. Col. (2): Target name. Cols. (3) and (4): Right Ascension and Declination coordinates of the source. Col. (5): Redshift of the source. Col. (6): Integration time of observations in units of hours. Col. (7): r-band magnitude. Col. (8): Morphology type based on SDSS. Col. (9): Sources with detection of radio jets (Y) and no jets (N).}
\label{table:1}
\end{deluxetable*}

\section{Measurements and Analyses}\label{section:analysis}

In this section, we discuss the main goals of our project. Additionally, we present our analysis recipes, including measuring SFRs and ionized and molecular gas outflow properties.

\subsection{Main Goals of the Project}

Our goals are to focus on the study of AGN feedback in nearby AGNs with and without powerful relativistic jets. We aim to accurately determine SFRs using the observed sub-mm JCMT/SCUBA-2 data and investigate the interplay between the activity levels of AGNs and/or jets and SFRs within their host galaxies. In our project, our sample contains AGNs with a broad range of luminosity with and without jets ($\mathrm{z}$ $<$ 0.3) exhibiting strong outflow signatures in their emission line profiles, characterized by velocity shifts of up to 400 \kms and dispersions reaching 600 \kms. This unique sample provides us with a great opportunity to investigate various modes of AGN feedback, including QSO- and radio-mode feedback. Particularly, we aim to investigate the enhancement or suppression of SFRs in these two subsamples to look for evidence of any connection between gas outflows and/or jets and star formation activities.


\subsection{Measuring Star-Formation Rates} 

\begin{figure*}
\centering
\includegraphics[width = 0.49\textwidth]{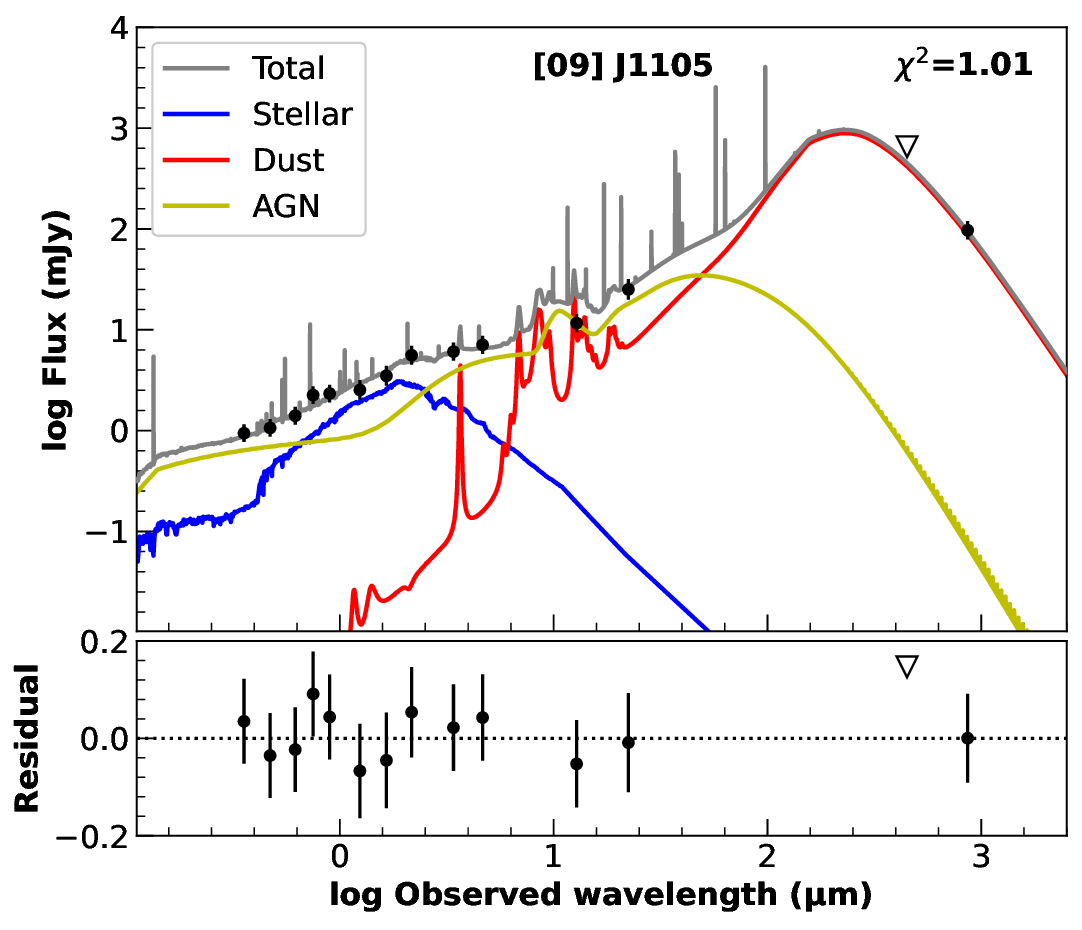}
\includegraphics[width = 0.49\textwidth]{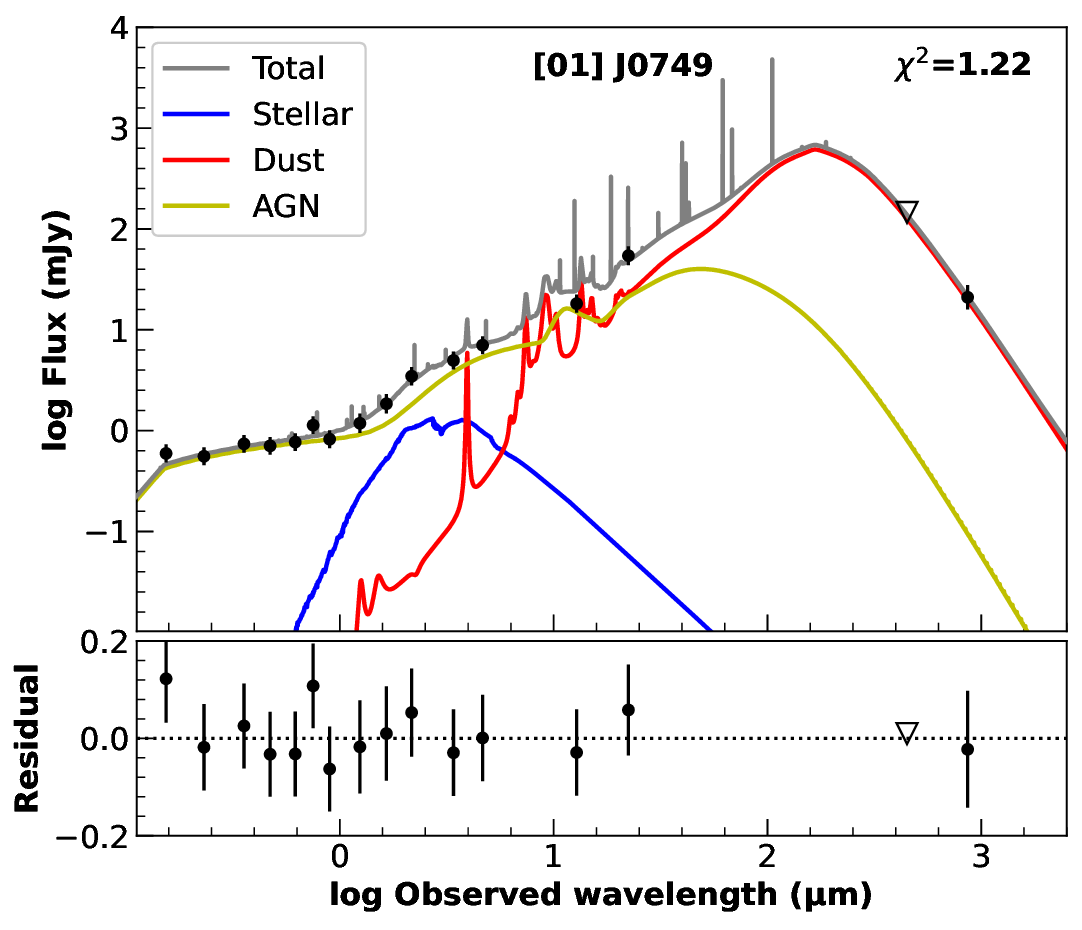}
\caption{SED fitting results of two targets in our proposals. The text in the figures provides explanations for the models and data points. For both of these targets, the SEDs, ranging from ultraviolet to mid-infrared bands, are dominated by AGN emission, while the far-infrared/sub-millimeter SEDs primarily result from dust re-emission. Therefore, the utilization of SCUBA-2 is crucial for accurately estimating the SFRs of our targets. The triangle indicates the upper limit of detection. \label{fig:SEDfitting}}
\end{figure*}

The sub-mm continuum fluxes at 450 $\rm \mu m$ and 850 $\rm \mu m$ play a crucial role in obtaining robust measurements of SFRs. We compile all available data in the literature for our sample, spanning from UV/optical to radio data. Subsequently, we conduct SED fitting, using tools such as CIGALE V2022, to estimate the total IR luminosity and estimate SFRs. CIGALE utilizes a Bayesian approach for analyzing the parameter space grids of SEDs, providing accurate estimations of values along with their associated uncertainties. To determine the SFRs, we depend on the Bayesian-derived values and their corresponding uncertainties of the dust luminosity. In our sample, the error estimations of sub-mm continuum fluxes vary from 48$-$70 mJy at 450 $\rm \mu m$ and 1$-$4 mJy at 850 $\rm \mu m$, respectively. The average SFR error of our sample is approximately 0.1 dex. A full analysis for SED fitting of our sample will be presented in \citet{Qin+23a}. Figure \ref{fig:SEDfitting} presents examples of our SED fitting for two targets in the program M22BP038. Similar to \citet{Kim+22}, from the obtained FIR flux, we determine the SFR by adopting the equation proposed by \citet{Kennicutt+98} as revised for the Kroupa IMF,  
\begin{equation}\label{eq:sfr}
\rm \log SFR_{\text{FIR,SED}}\ (\text{M}_\odot \, \text{yr}^{-1}) = \rm \log L_{\text{FIR}}\ (\text{erg} \, \text{s}^{-1}) - 43.519,
\end{equation}
Here, $\rm L_{FIR}$ refers as the total IR luminosity over 8 $-$ 1,000 $\rm \mu m$. 

\begin{figure}
\centering
\includegraphics[width = 0.45\textwidth]{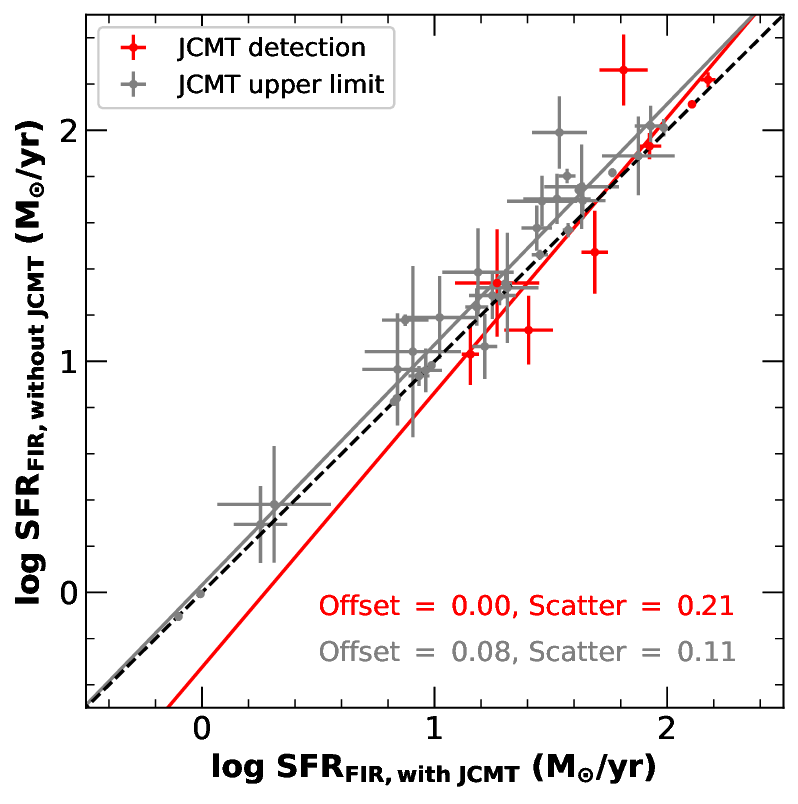}
\caption{Comparation of SFRs determined from SED fittings with and without including JCMT data. The red and gray solid lines show the best fits for sources with detected and upper limit detections in the JCMT data, respectively. The dashed line displays the one-to-one relation. \label{fig:compare_sedsfr}}
\end{figure}

To examine the roles of sub-mm continuum fluxes at 450 $\rm \mu m$ and 850 $\rm \mu m$ in determining SFRs, we perform SED fitting with and without using JCMT data. In Figure \ref{fig:compare_sedsfr}, we present a comparison of SFRs derived from dust luminosities using SED fitting, with and without incorporating JCMT data, highlighting the offset and scatter between these two sets of results. Our analysis reveals that sources with only JCMT upper limits (gray dots) tend to yield overestimated SFRs with an offset of 0.08 dex, a scatter of 0.11 dex, and a best fit slope of 1.041 when JCMT data are omitted. Furthermore, sources detected by JCMT exhibit a significant scatter of 0.21 dex, with a steeper slope of 1.189 compared to a one-to-one relation. Notably, certain sources, well-detected by IRAS and Herschel, provide robust FIR data, exemplified by four points with minimal uncertainty in the lower left and upper right corners of Figure \ref{fig:compare_sedsfr}. This underscores the critical role of FIR and sub-mm data in the CIGALE fitting model for reducing uncertainties in SFR determinations. 


\subsection{Gas Outflow Properties} \label{section:gasoutflow}

Our analyses include the utilization of fitting routines established in our previous works \citep[e.g.,][]{Le+17b, Le+19, Le+20, Le+23} to fit the observed spectra and measure emission-line properties for both DBSP and TPSP. The high-quality long-slit spectra obtained from TPSP and DBSP are invaluable for accurate spatially resolved measurements of the observed emission lines ranging from UV-optical to NIR bands. 

\begin{figure*}
\centering
\includegraphics[width = 0.90\textwidth]{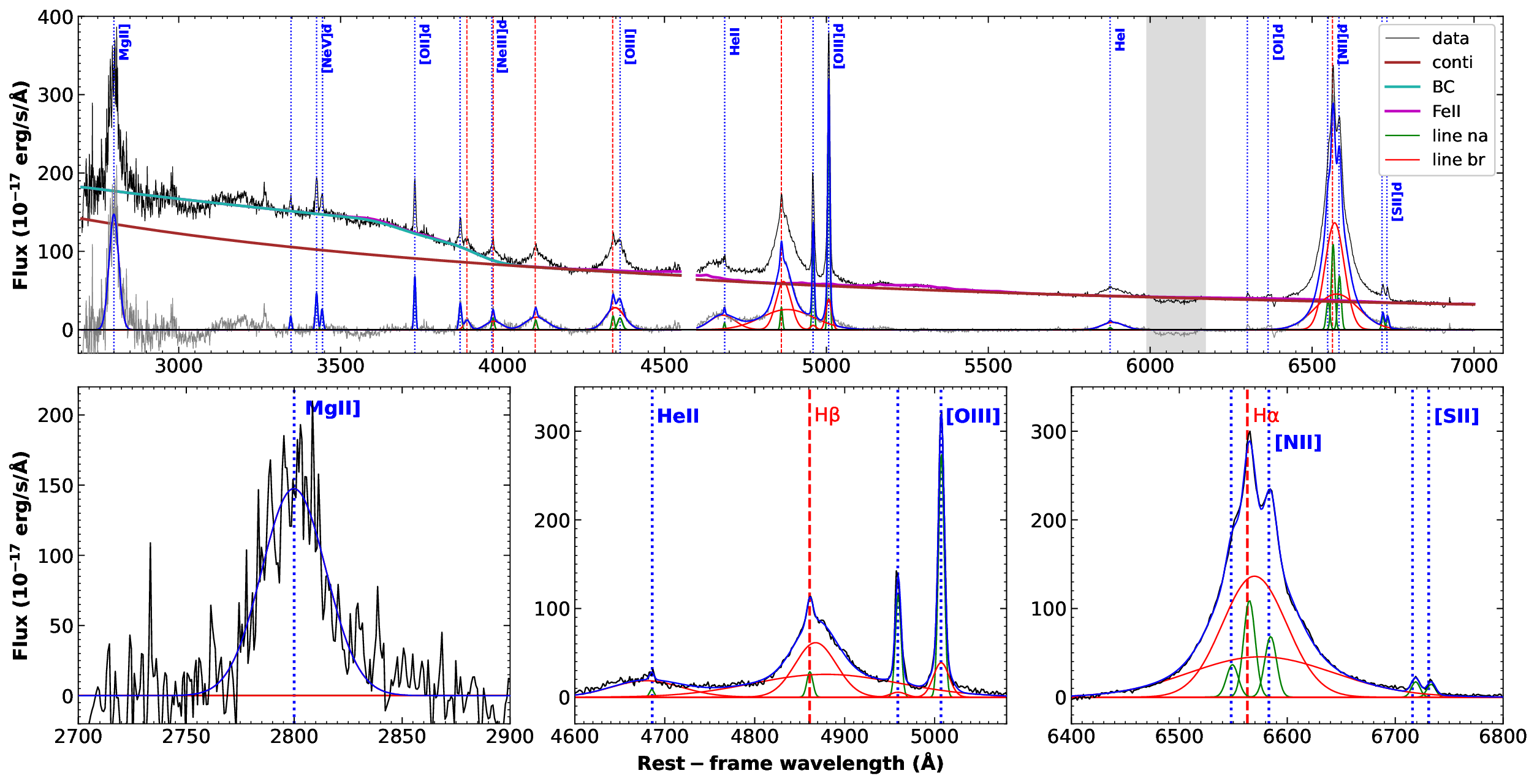}
\caption{An integrated DBSP spectrum of J0749+4510 ($z=0.192$, $\rm m_{r}=16.78$ mag). We fitted the spectrum using the BADASS package \citep{Sexton+20}. Top panel: the observed spectrum (black) and fitting models, including AGN power-law continuum (purple), Balmer-continuum (cyan), Fe II emission (magenta), narrow (green), and broad (pink) emission line models. The shaded region around the 6100 \AA\ continuum is discarded from the spectral fitting. Bottom panels: the fitting models of emission lines show multiple broad and narrow components in their profiles, e.g., Mg II, H$\beta$, [O III], H$\alpha$. The high S/N spectrum with detected multiple emission lines gives us a great interest in exploring the ionized gas kinematics along the long-slit.\label{fig:pyqsofit}}
\end{figure*}

\begin{figure*}
\centering
\includegraphics[width = 0.92\textwidth]{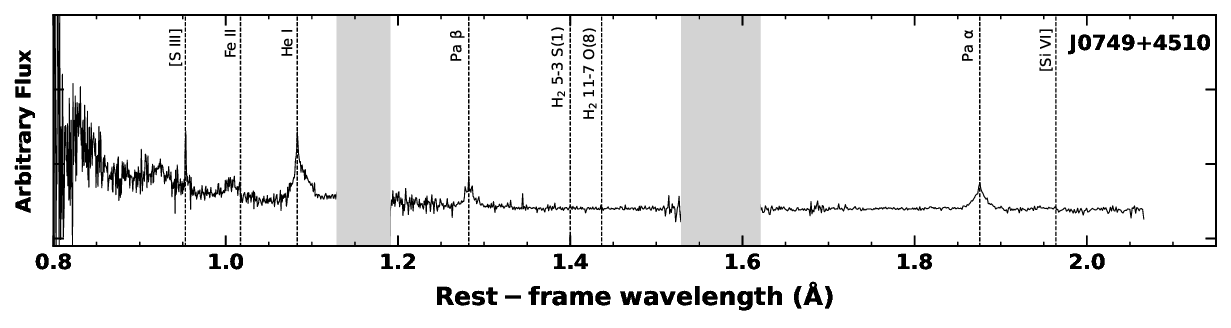}
\caption{An integrated TPSP spectrum of J0749+4510 ($z=0.192$, $\rm m_{r}=16.78$ mag). NIR emission lines, e.g., [S III], He I, $\rm P\beta$, $\rm P\alpha$, and $\rm H_{2}$ are indicated by vertical dashed lines. The shaded regions denote areas discarded from the spectrum due to significant noise influence from skylines or edge regions of the TPSP spectrograph. \label{fig:triplespec}}
\end{figure*}

Figure \ref{fig:pyqsofit} shows an example of an integrated spectrum of J0749+4510 observed by the DBSP by using Bayesian AGN Decomposition Analysis for SDSS spectra (BADASS, \citealp{Sexton+20}). BADASS is a comprehensive tool designed for the simultaneous fitting of multiple spectral components, encompassing the power-law continuum and stellar emission, as well as both the narrow and broad components of AGN emission lines. The fitting procedure employs Markov chain Monte Carlo techniques, ensuring the derivation of reliable uncertainties for the fitted parameters. In our sample, the error measurements are approximately 0.1 dex for continuum luminosities and line kinematics and, consequently, about 0.3 dex for black hole masses. The best fitting models of emission lines show multiple broad and narrow components in their profiles, e.g., Mg II, H$\beta$, [O III], H$\alpha$, [N II] lines. Figure \ref{fig:triplespec} displays an integrated spectrum of J0749+451 observed by TPSP. We are able to detect multiple emission lines in the NIR band, e.g.,  [S III] 0.9533 $\rm \mu m$, He I 1.0832 $\rm \mu m$, $\rm P\beta$ 1.2822 $\rm \mu m$, $\rm P\alpha$ 1.8756 $\rm \mu m$, and obtain possible detections of various molecular lines, such as rovibrational $\rm H_{2}$ emission lines \citep[e.g.,][]{Le+17a}. 

From the observed DBSP and TPSP spectra, we aim to measure outflow properties, such as energies and kinematics, and then properly compare these properties between observed emission lines in UV-optical and NIR bands. We can also compare the outflow sizes between the molecular and ionized gas components. We calculate the outflow size based on molecular gas, such as $\rm H_{2}$, and compare it with the size of the outflows measured using ionized gas, like [O III]. Furthermore, we derive mass outflow rates and kinetic power for the molecular outflows and compare them with those of the ionized gas outflows. Detailed analyses of kinematic comparisons between the DBSP and TPSP spectra will be presented in \citet{Xia+23}. 

\subsection{Spatially Resolved Gas Kinematics} \label{section:spatial}

\begin{figure*}
\centering
\includegraphics[width = 0.7\textwidth]{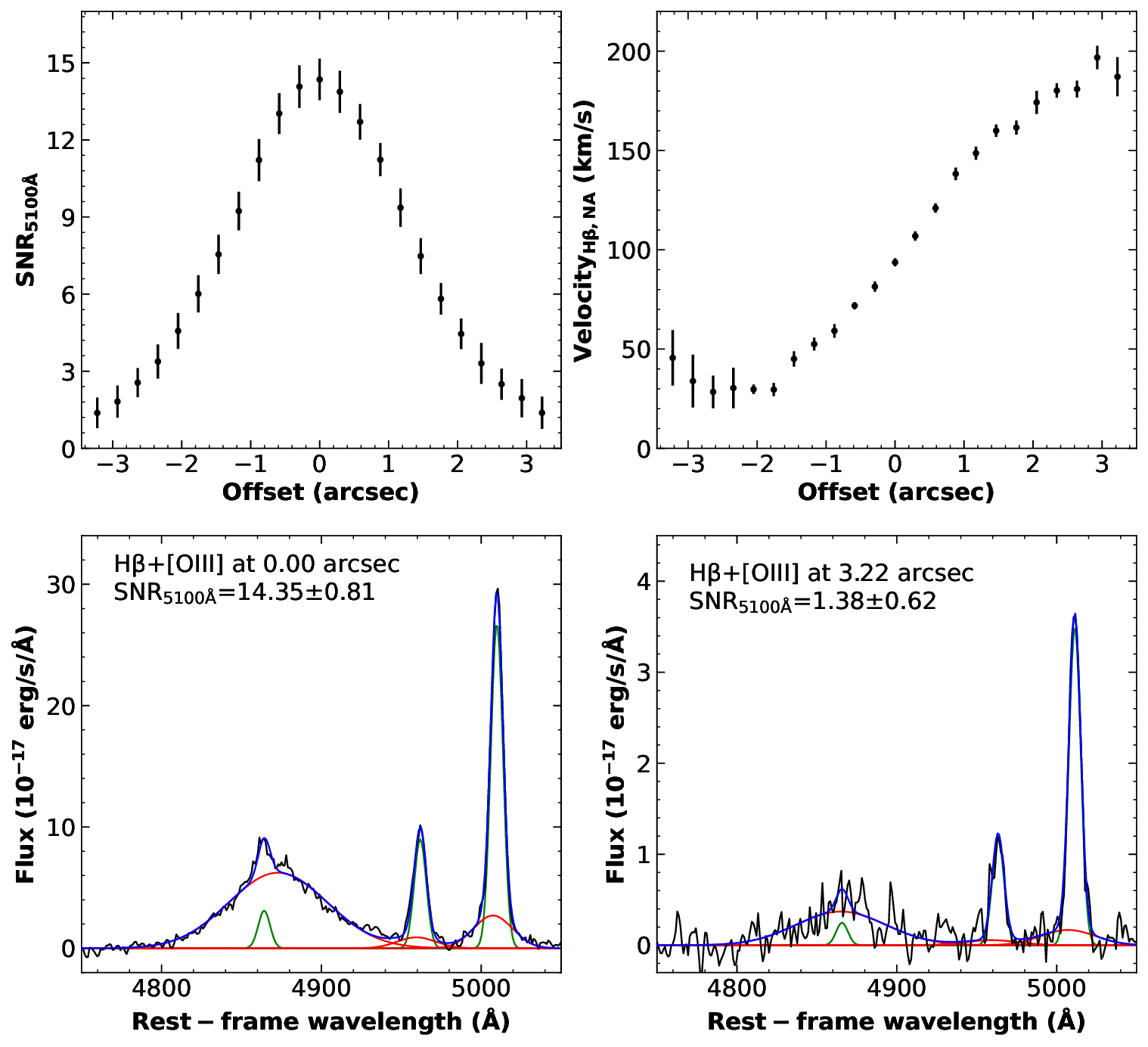}
\caption{Spatially resolved spectra of J0749+4510 ($z=0.192$, $\rm m_{r}=16.78$ mag) obtained with DBSP. In the top left panel, we present the S/N at the 5100\AA\ continuum as a function of position along the slit direction. On the top right, we show the measured velocities of the H$\beta$ narrow component vary along the slit direction. The bottom panels display the extracted spectra of the H$\beta$ and [O III] emission lines at 4969\AA\ and 5007\AA\ for both the central slit region (positioned at 0") and the outer edge slit region (positioned at 3.22") in the left and right panels, respectively. \label{fig:spatial_spec}}
\end{figure*}

The high S/N spectra with detected multiple emission lines give us a great interest in exploring the gas kinematics along the long-slit. We measure the kinematics of the emission lines, including velocity shift and velocity dispersion, using single or multiple Gaussian functions. Specifically, we aim to extract spectra from spatial bins along the length of the slit, each approximately 1" to 2" in size. Figure \ref{fig:spatial_spec} presents the resolved kinematic measurements of the H$\beta$ emission line along the major axis of J0749+4510 observed by the DBSP. The spatially resolved spectra show high S/N at the 5100\AA\ continuum, S/N $\sim$ 15 at the central slit region, and S/N $\sim$ 6 at the outer edge slit region (positioned at 2"), respectively. The spatially resolved velocities of the H$\beta$ narrow component show clear rotation with the maximum projected velocity of $\sim$ 200 \kms. Particularly, we can see that there is a difference in the rotation curve velocities of the H$\beta$ narrow line between the minus and positive offsets along the slit. This discrepancy could be relative to the launching jet aligning with the positive offset direction, as shown in the radio image of this target. 

The analysis of the resolved kinematics helps us understand how the kinematics of emission lines vary as a function of distance from the core to the outer regions of the AGN structure. We aim to use spatially resolved emission lines from the full wavelength range spectra to calculate emission line flux ratios. These ratios quantify the radiative influence of the AGN as a function of distance along the slit. Consequently, we can constrain the dust opacity within the nuclear region and host galaxy, which could significantly affect the observational characteristics of gas outflows. Lastly, we investigate gas outflows at larger distances from the AGN, using lower ionization potential emission lines, such as H$\alpha$ and $\rm H_{2}$ emission lines. Furthermore, we can estimate SFRs based on [O II] and H$\alpha$ emissions \citep{Zhuang+19}. We can have a proper comparison between SFRs determined from UV-optical emission lines and SFRs obtained from JCMT data. This allows us to have a comprehensive study of SFRs based on multiple tracers and quantify the discrepancies among them. A study of spatially resolved spectra for comparing the outflow energies, kinematics, and SFRs could provide valuable insights into the physical nature of the relationship between AGN strength and SF activities, such as inside-out growth. The detailed analyses and results of this study will be presented in \citet{Qin+23b}.

\subsection{Connection between Gas Outflows and Radio Jets?} \label{section:gasjet}

Some studies have suggested an implicit connection between ionized gas outflows and large-scale radio jets \citep[e.g.,][]{Tadhunter+14, Le+17b, Ayubinia+22}. The deep and high-quality spectra obtained from DBSP and TPSP are invaluable for achieving accurate spatially resolved measurements of the observed emission lines. These measurements enable us to investigate the relationship between the radio jet and gas outflows. Leveraging the high-spatial-resolution of outflow gas kinematics, along with robust measurements of SFRs in our project, offers an excellent opportunity to study radio mode feedback.

Our initial sample for this purpose comprises 5 AGNs, each exhibiting very high radio luminosity as observed by VLA FIRST at 1.4 GHz ($\rm L_{1.4 GHz} > 10^{41} erg\ s^{-1}$). These AGNs display strong outflow signatures in the ionized gas kinematics, with velocity shifts of up to 400 $\rm km\ s^{-1}$ and dispersions reaching 600 $\rm km\ s^{-1}$. These 5 nearby AGNs are exceptionally unique, allowing us to explore the connection between outflow kinematics and radio activity. The long-slit spectroscopic observations from TPSP and DBSP provide us with spatially-resolved measurements spanning several kiloparsecs along the respective slit directions. This allows us to probe the signatures of the connection between gas outflows and jets. Furthermore, we have observed our targets using slits at different position angles (PA) to cover the entire emission gas with varying geometric distributions within AGN systems. The diverse PA slit observations with respect to the jet axis enable us to compare ionized gas kinematics across different geometric distributions in AGN systems, enhancing our understanding of the relationship between ionized gas outflows and radio jets.


Our goal is to extract spectra from spatial bins along the length of the slit, ranging from 1" to 2", as we have shown in Section \ref{section:spatial}. This approach allows us to investigate how the kinematics of emission lines, such as [O III] and $\rm H_{2}$, vary as a function of distance from the core and along the radio jet. We can conduct comparisons between outflow energies, kinematics, and SFRs along the extended jet on scales spanning several kiloparsecs. Additionally, we compare the outflow gas kinematics and SFRs of sources without jets to identify potential differences in outflow energies, kinematics, and SFRs between AGNs with and without jets. Furthermore, we determine the black hole mass of our sample based on our improved single-epoch black-hole mass, which employs H$\beta$ emission \citep{Le+20}. This provides us with reliable estimates of black hole mass and, consequently, black hole accretion rate and Eddington ratio. The detailed analysis and the subsequent results of this study will be presented in \citep{Qin+23b}.


\section{Initial Results and Future Plans}\label{section:result}


\subsection{Initial Results}

Recently, \citet{Kim+22} provided SFRs measured with SCUBA-2 for 52 AGNs ($\mathrm{z}$ $<$ 0.2) and found a strong correlation between SFRs and Eddington ratios. The results of \citet{Kim+22} underscore the importance of FIR and/or SCUBA-2 data in accurately estimating SFRs. In \citet{Kim+22}, they focused on the relation between outflow gas properties and SFRs. In our project, our sample includes nearby AGNs ($\mathrm{z}$ $<$ 0.3 ) with and without powerful relativistic jets. We aim to study if there is any connection between gas outflows and/or jets and SFRs. Such a study can reveal the roles of AGN outflows and/or jets in enhancing or quenching SF in their host galaxies.

The unique observed sub-mm JCMT data are important to obtain robust SFRs for our sample. Also, using the long-slit spatially resolved spectra from TPSP and DBSP, as well as integrated spectra from SDSS, we study in detail the connections between multiphase gas outflows, jets, and SFRs. The results of our project, complemented with the results from \citet{Kim+22}, could deepen our understanding of the interplay between AGNs/jets and star formation activities. 

\begin{figure}
\centering
\includegraphics[width = 0.48\textwidth]{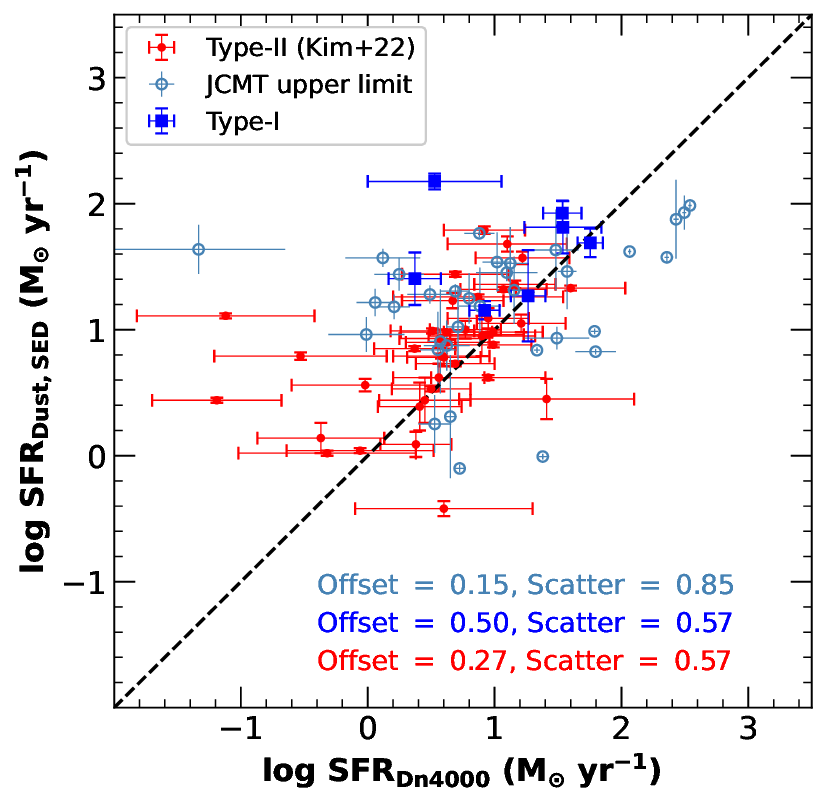}
\caption{Comparison of SFRs, estimated through SED fitting versus $\rm SFR_{Dn4000}$ for targets in our proposal. $\rm SFR_{Dn4000}$ is estimated using the Dn4000\AA\ values from the MPA-JHU catalog \citep{Brinchmann+04}. Additionally, we compare our results with a sample of Type-2 AGNs \citep{Kim+22}, estimating SFRs through SED fitting, including continuum fluxes observed at 450 $\rm \mu m$ and 850 $\rm \mu m$ from SCUBA-2. Offset refers to the mean difference between the values on the y-axis and the corresponding values on the x-axis, while scatter represents the standard deviation of these differences. \label{fig:sfr_compare}}
\end{figure}

In Figure \ref{fig:sfr_compare}, we present a comparison of the determined SFRs using Equation \ref{eq:sfr} for some sources in our initial sample, employing SED fitting, including continuum fluxes observed at 450 $\rm \mu m$ and 850 $\rm \mu m$ from SCUBA-2, with SFRs traced by the 4000\AA\ break, as estimated by the Max Planck Institute for Astrophysics and Johns Hopkins University (MPA-JHU) catalog \citep{Brinchmann+04}. Additionally, we include the results from previous work by \citet{Kim+22} for reference and comparison with our findings. SFRs determined by the SED fitting tend to be higher than those derived from the 4000\AA\ break ($\rm SFR_{Dn4000}$). This discrepancy could be attributed to potential uncertainties associated with AGNs in the MPA-JHU catalog. In our recent study \citep{Le+24a}, we found that the sample used in the MPA-JHU catalog primarily consists of young galaxies, leading to significant scattering and uncertainties when applied to older and redder galaxies (e.g., $\mathrm{D_{n}4000}$ > 1.8). 

Additionally, there are also some limitations of SFR estimation in our sample that we need to consider. Using CIGALE fitting for SFR estimation may present challenges due to its complex modeling of galaxy SEDs, potential degeneracies in parameter constraints, and sensitivity to dust modeling. The process involves intricate modeling of the entire spectral energy distribution, which can lead to uncertainties and biases in derived SFRs. Also, uncertainties in flux calibration of sub-mm data can propagate into the SFR estimates, particularly for targets with sub-mm upper limits.

Moreover, our selected sample contains targets that show strong outflow gas signatures with powerful relativistic jets. Powerful radio jet emission can also contribute to the observed sub-mm bands, particularly through synchrotron emission \citep[e.g.,][]{Rojas+21}. This additional emission component needs to be distinguished from the observed sub-mm fluxes to avoid overestimating SFRs. We utilize radio data and best fit models provided by SPECFIND V3.0 \citep{Stein+21} for our sample to predict the contribution of radio jet emission to sub-mm data. We estimate the ratio between observed sub-mm fluxes and the predicted fluxes of the jet at 450 $\rm \mu m$ and 850 $\rm \mu m$, respectively. If the linear ratios are smaller than 3, we identify sub-mm fluxes as upper limits. In contrast, if the ratios are larger than 3, we subtract the predicted jet emission values from the observed sub-mm fluxes. The detailed analysis and the subsequent results of this study will be presented in \citet{Qin+23a}.

In short, there is no perfect SFR indicator; each will have its own limitations, as discussed in Section \ref{sec:intro}. In this project, our focus is on the FIR and sub-mm continuum flux diagnostic, which primarily tracks cold dust emission resulting from star formation. These indicators are crucial for obtaining robust SFR estimates from the SED fitting model. Increasing the sample size with FIR/sub-mm flux measurements is essential for gaining insights into star formation processes and comprehending the physical properties of AGNs.

\subsection{Future Plans} \label{section:pfs}

\begin{figure}
\centering
\includegraphics[width = 0.48\textwidth]{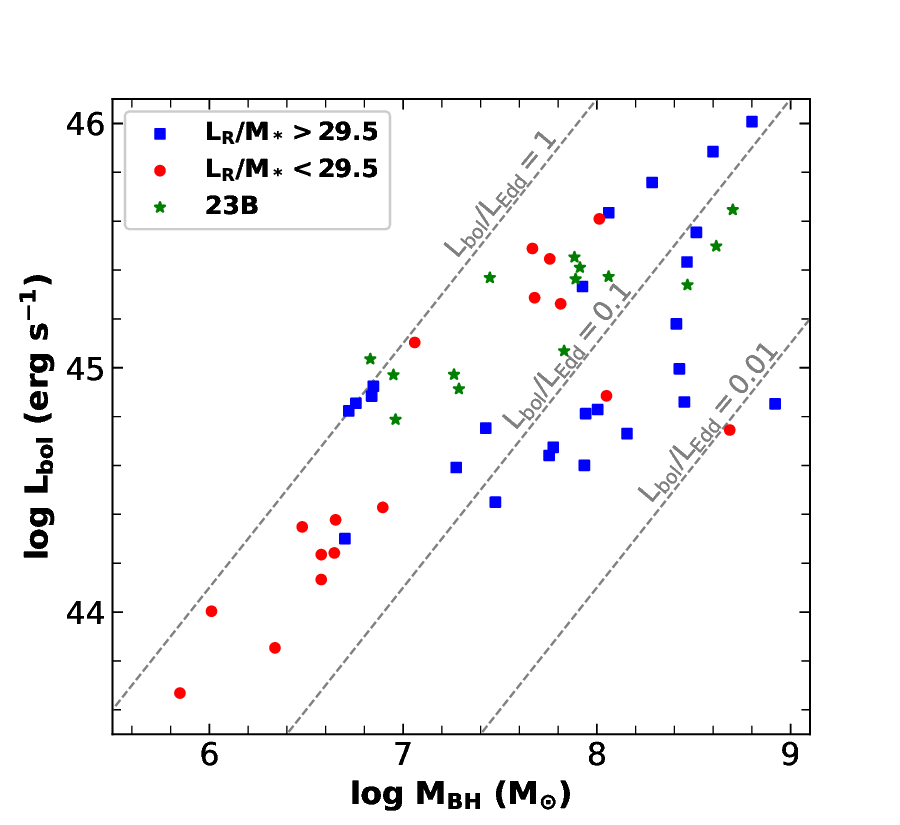}
\caption{Distributions of the black hole masses and bolometric luminosities of our selected sources. The sample of the proposals 2022B and 2023A semesters (M22BP038 and M23AP023) are shown in blue ($\rm L_{1.4 GHz}$/$\rm M_{*}$ $>$ 29.5) and red ($\rm L_{1.4 GHz}$/$\rm M_{*}$ $<$ 29.5) colors. Green stars show the sources of the proposal 2023B semester (M23BP013). Dash lines indicate different Eddington ratios.\label{fig:bol_agn_lum}}
\end{figure}

Our sample spans a broad luminosity range for AGNs with jets, with bolometric luminosities ($\rm L_{Bol}$) and black hole masses ($\rm M_{BH}$) ranging from $10^{43}$ to $10^{46}$ erg s$^{-1}$ and $\rm \log M_{BH}/M_{\odot}$ $\sim$ 6$-$9, respectively (Figure \ref{fig:bol_agn_lum}). However, for the AGNs without jets, our targets mostly belong to the high luminosity ($\rm L_{Bol}$ $\gtrsim$ $10^{45}$ \ergs) and massive black holes ($\rm \log M_{BH}$/$\rm M_{\odot}$ $\gtrsim$ 7) category. To broaden our sample and encompass a wider range of luminosities, we aim to propose additional observations targeting lower-luminosity objects ($\rm L_{Bol}$ $\lesssim$ $10^{45}$ \ergs) and less massive black holes ($\rm \log M_{BH}$/$\rm M_{\odot}$ $\lesssim$ 7) using JCMT/SCUBA-2. Furthermore, while the majority of targets in our current sample are type-1 AGNs, we recognize the importance of including type-2 AGNs to ensure a comprehensive understanding. By incorporating observations of type-2 AGNs, we aim to provide a more balanced representation and enhance our understanding of the diverse population of AGNs. This comparative analysis of SFRs between type-1 and type-2 sources will enrich our dataset and help deepen our understanding of AGN activities and star-formation processes in nearby active galaxies.

It is also crucial to enlarge the sample in this study to higher-redshift galaxies (e.g., $\mathrm{z}$ $>$ 0.5). We plan to propose to observe higher-redshift galaxies using the JCMT and the PFS observations to study the coevolution dynamics between AGN activity and star formation. The PFS, which we can access, is currently under construction for the 8.2 m Subaru Telescope in Maunakea, Hawaii, and is scheduled to commence observations in early 2025. With 2394 fibers covering a 1.25 $\rm deg^{2}$ field of view, the PFS offers wide wavelength coverage from UV to optical to NIR (0.38 $-$ 1.25 $\rm \mu m$). This instrument holds great promise for investigating the physical properties and evolution of galaxies and AGNs, especially for high-redshift objects ($\mathrm{z}$ $>$ 0.5) in the coming decades. Offering significant advancements over previous spectroscopic surveys in terms of depth, sample size, spectral resolution, S/N, and object density, the PFS will play a pivotal role in addressing numerous fundamental questions in the field of AGNs, notably on the formation and evolution of galaxies. As part of the observing strategy and target selection plans, the PFS will prioritize selecting 10,000 sources from the Chandra and XMM-Newton surveys and 500 sources from the SCUBA-2 and FIRST surveys for studying galaxy evolution \citep{Greene+22}.

\section{Summary}\label{section:sum}

In this paper, we introduce our project, AGNSTRONG, which aims to investigate the interplay between the activity of galactic nuclei and star formation processes in their host galaxies. Our primary approach involves improving the accuracy of SFR estimation using sub-mm data obtained from the JCMT/SCUBA-2. Our sample comprises 21 AGNs with jets and 14 AGNs without jets, all of which have been observed or scheduled for observations by SCUBA-2. Our main focus within this sample is the precise determination of SFRs and the subsequent exploration of the relationships between AGN activity and SFRs. Our project encompasses the study of AGN feedback, including both QSO- and radio-mode feedback. We also utilize UV/optical and NIR spectroscopic observational data from DBSP and TPSP at the P200 telescope to examine the spatially resolved spectra of selected sources within our sample. We aim to conduct an in-depth analysis to evaluate whether there is an increase or suppression of SFRs in targets with and without powerful relativistic jets. Here are the current status and future plans of our AGNSTRONG project:

\smallskip
(1) All 21 sources with jets have completed observations using JCMT/SCUBA-2, while observations for the remaining 14 sources without jets are ongoing. Combining sub-mm data with FIR data is essential for estimating robust SFRs through SED fitting. We intend to propose observations for additional targets, including both type-1 and type-2 AGNs, in the upcoming proposal calls.

\smallskip
(2) Within our sample, 11 out of 35 sources have been observed with DBSP and TPSP. The high-quality spectra obtained from these instruments are invaluable for accurately measuring spatially resolved emission lines. This will enable us to make comprehensive comparisons of outflow energies, kinematics, and SFRs, providing valuable insights into the physical relationship between AGN activity and star formation in AGNs with and without jets.

\smallskip
(3) Our future plans involve extending our sample to higher-redshift targets. Data from the PFS project will provide an excellent opportunity to study the relationship between AGN activity and star formation in higher-redshift active galaxies.

\bigskip

We thank the referee for valuable suggestions and comments that improved the presentation and clarity of this paper. This work has been supported by the National Natural Science Foundation of China (NSFC-12003031, NSFC-12203047, NSFC-12393814, NSFC-12025303). H. A. N. Le acknowledges the support from the Fundamental Research Funds for the Central Universities (WK2030000057). We thank the kind support of the `Exchange Partner Program' of The James Clerk Maxwell Telescope. The SCUBA-2 data in this paper were obtained under the program IDs: M22BP038, M23AP023, and M23BP013. The James Clerk Maxwell Telescope is operated by the East Asian Observatory on behalf of The National Astronomical Observatory of Japan; Academia Sinica Institute of Astronomy and Astrophysics; the Korea Astronomy and Space Science Institute; the National Astronomical Research Institute of Thailand; Center for Astronomical Mega-Science (as well as the National Key R\&D Program of China with No. 2017YFA0402700). Additional funding support is provided by the Science and Technology Facilities Council of the United Kingdom and participating universities and organizations in the United Kingdom and Canada. Additional funds for the construction of SCUBA-2 were provided by the Canada Foundation for Innovation. We wish to recognize and acknowledge the very significant cultural role and reverence that the summit of Maunakea has always had within the indigenous Hawaiian community. We are most fortunate to have the opportunity to conduct observations from this mountain. This research uses data obtained through the Telescope Access Program (TAP), which has been funded by the TAP member institutes (Proposal ID: CTAP2022-A0037). We thank the kind help to obtain the TPSP and DBSP data in this paper from Yibo Wang, Zhenfeng Sheng, and Mengqiu Huang.

\end{CJK*}

\end{document}